\documentclass[runningheads,a4paper]{llncs}
\usepackage[left=4.4cm,right=4.4cm,top=5.2cm,bottom=5.2cm]{geometry}
\usepackage{doi}

\usepackage{amssymb}
\usepackage{graphicx}
\usepackage{color}

\usepackage{mathtools,bm}

\usepackage[
backend=biber,
bibencoding=utf8,
firstinits=true,
url=false,
isbn=false,
]{biblatex}
\AtEveryBibitem{\clearlist{language}}
\AtEveryBibitem{\clearfield{month}}

\DeclareNameAlias{default}{last-first}

\renewbibmacro*{doi+eprint+url}{%
    \printfield{doi}%
    \newunit\newblock%
    \iftoggle{bbx:eprint}{%
        \usebibmacro{eprint}%
    }{}%
    \newunit\newblock%
    \iffieldundef{doi}{%
        \usebibmacro{url}}%
        {}%
}

\newcommand{\keywords}[1]{\par\addvspace\baselineskip
\noindent\keywordname\enspace\ignorespaces#1}

\newcommand{\ignore}[1]{}
\begin{document}

\mainmatter  

\title{Sparse Coding Predicts Optic Flow Specificities of Zebrafish Pretectal Neurons}

\titlerunning{Learning Optic Flow Specificities of Pretectal Neurons}

%
%
\author{Gerrit A.~Ecke \and Fabian A.~Mikulasch \and Sebastian A.~Bruijns
  \and Thede Witschel \and \\ Aristides B.~Arrenberg \and Hanspeter
  A.~Mallot}
\authorrunning{Ecke et al.}

\institute{Dept.\ of Biology, University of T\"ubingen, T\"ubingen, Germany \\
gerrit.ecke@uni-tuebingen.de \\
hanspeter.mallot@uni-tuebingen.de }

%
%

\maketitle

\begin{abstract}
  Zebrafish pretectal neurons exhibit specificities for large-field
  optic flow patterns associated with rotatory or translatory body
  motion. We investigate the hypothesis that these specificities
  reflect the input statistics of natural optic flow. Realistic motion
  sequences were generated using computer graphics simulating
  self-motion in an underwater scene. Local retinal motion was
  estimated with a motion detector and encoded in four populations of
  directionally tuned retinal ganglion cells, represented as two
  signed input variables. This activity was then used as input into
  one of two learning networks: a sparse coding network (competitive
  learning) and backpropagation network (supervised learning). Both
  simulations develop specificities for optic flow which are
  comparable to those found in a neurophysiological study~\cite{KuboETAL14}, 
  and relative frequencies of the various neuronal
  responses are best modeled by the sparse coding approach. We
  conclude that the optic flow neurons in the zebrafish pretectum do
  reflect the optic flow statistics. The predicted vectorial receptive
  fields show typical optic flow fields but also ``Gabor'' and
  dipole-shaped patterns that likely reflect difference fields needed
  for reconstruction by linear superposition.

\end{abstract}

\section{Introduction} 

\subsubsection*{Optimality of visual receptive fields.}

In his {\em ``neuron-doctrine for perceptual psychology'',} Horace
Barlow~\cite{Barlow72} suggests that the {\em ``nervous system is
  organized to achieve as complete a representation of the sensory
  stimulus as possible with the minimum number of active neurons''.}
This idea also underlies a number of theoretical approaches to visual
processing, such as independent component analysis, 
sparse coding, 
predictive coding, 
etc.;
for an overview see~\cite{HyvHurHoy09}. While the general approach
is widely accepted, specific predictions about the optimal processing
scheme will depend on the choice of the optimality criterion employed
as well as on the information requirements of each species'
life-style. Empirical tests of optimal coding theories of visual
processing are therefore often limited to a qualitative level. 

For the case of mammalian V1 cortex, Olshausen and Field~\cite{OlsFie05} have
summarized the evidence and concluded that for a full understanding of
the system, simultaneous measurements of the activities of a large,
unbiased set of neurons in response to natural stimuli would be
required. Two-photon calcium imaging allows to record activity from large 
populations of neurons. In {\em Drosophila}, simultaneous 
monitoring of more than 100 cells from the mushroom body has proven robustly sparse, but 
non-localized responses to varieties of odors~\cite{HonCamTur11}. Insights 
into functional aspects of memory and learning have been gained that extend
findings from single cell recordings
which show that sparsity is implemented by means of a normalizing feedback loop on a cellular 
level~\cite{PapCasNowLau11}.

\begin{figure}[t]
\centering
\setlength{\unitlength}{0.1\textwidth}
\begin{picture}(9.8,3)
\put(0.0,0){\resizebox{3\unitlength}{!}{\includegraphics{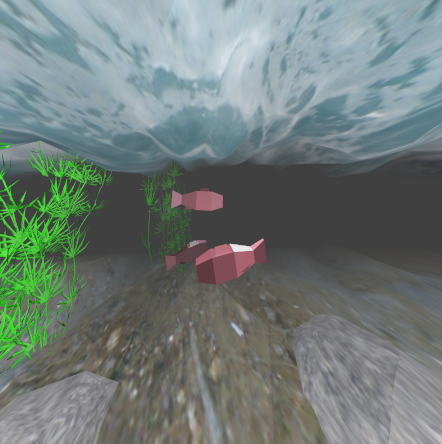}}}
\put(0.1,0.1){\makebox(0,0)[bl]{\textcolor{white}{\bf a.}}}

\put(3.4,0){\resizebox{3\unitlength}{!}{\includegraphics{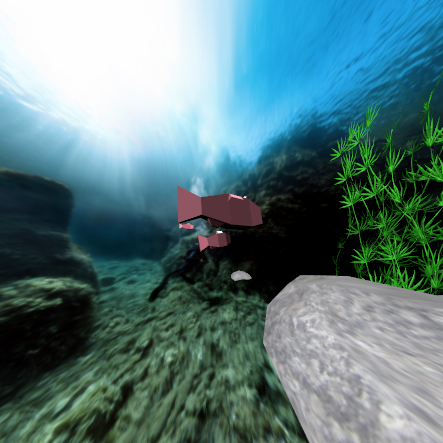}}}
\put(3.5,0.1){\makebox(0,0)[bl]{\textcolor{white}{\bf b.}}}

\put(6.8,0.0){\resizebox{3\unitlength}{!}{\includegraphics{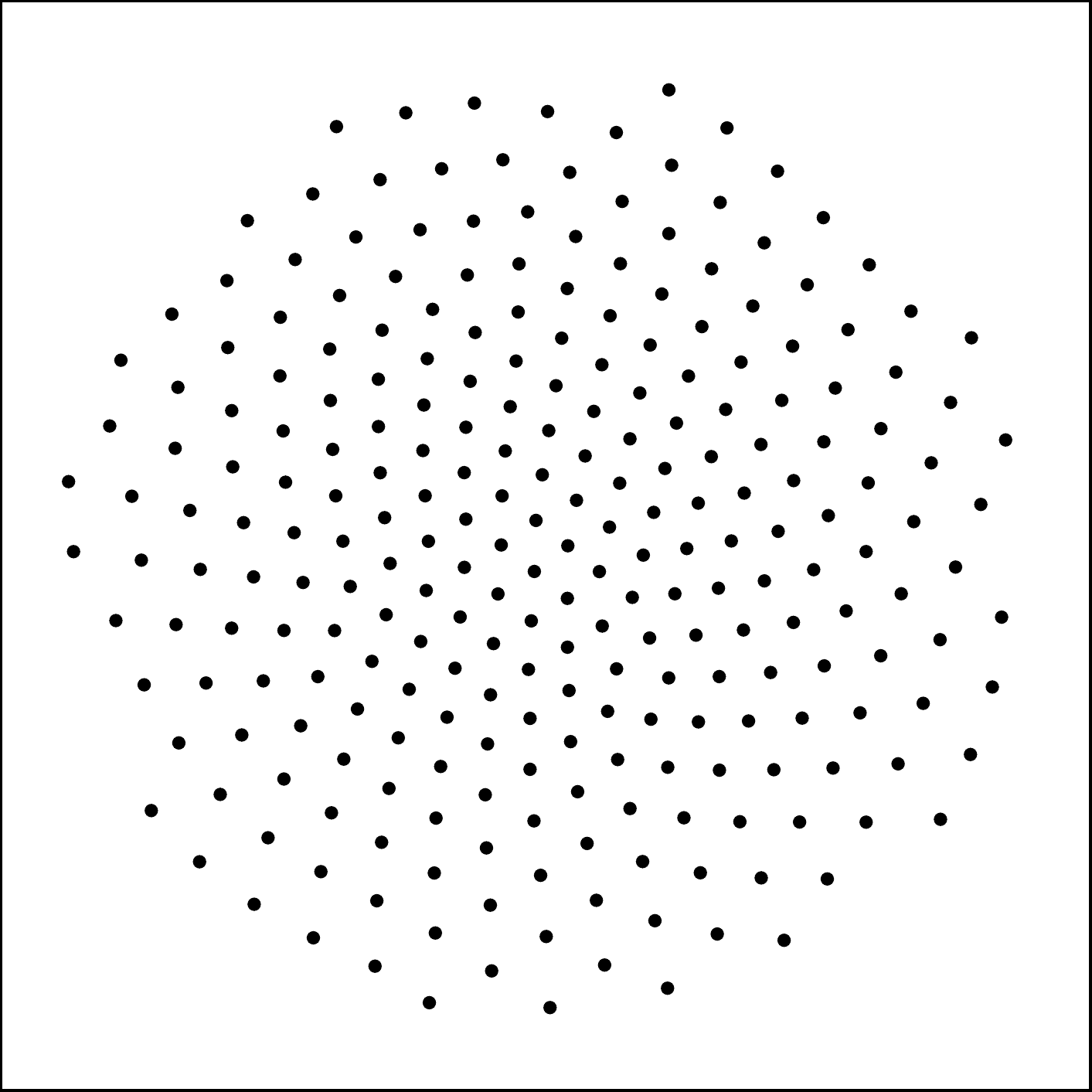}}}
\put(6.9,0.1){\makebox(0,0)[bl]{\textcolor{black}{\bf c.}}}

\end{picture}
\caption{{\bf a.} View of the virtual fish tank with muddy water (low
  viewing distance). Additional fish and plants will generate optic
  flow discontinuities. {\bf b.} Example with high
  visibility. {\bf c.}  Mosaic of retinal ganglion cells, used to
  calculate the motion input. }
\label{fig:sim_examples}
\end{figure}
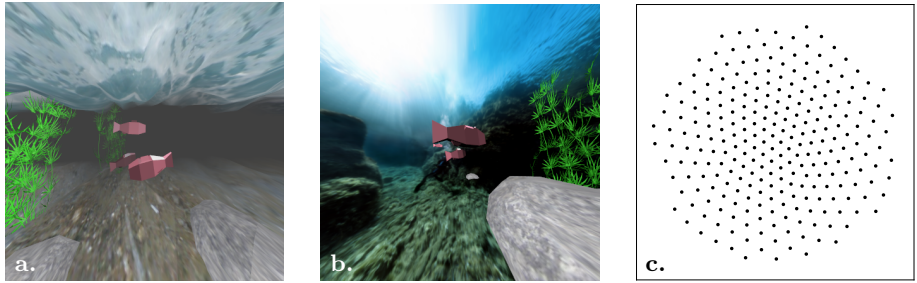

We attempt an analysis of this type for the area
pretectalis (APT) of the zebrafish, for which the response of thousands
of neurons has indeed been recorded while the fish is presented with
optic flow stimuli~\cite{KuboETAL14}. Experimentally found response properties 
from a large, representative sample of neurons will be compared to responses 
predicted from receptive fields of nodes in a artificial neural network trained 
with optic flow patterns that were generated by simulating observer movement in 
a virtual fish tank. The receptive field predictions will be based on two 
theoretical approaches, (\emph{i.}) sparse coding of optic flow patterns (unsupervised) and,
for comparison, (\emph{ii.}) 
backpropagation learning of ego-motion parameters from the same optic flow patterns (supervised).

\subsubsection*{Optic flow.}

Like many other animals, zebrafish larvae generate optokinetic
responses of the eyes (OKR) and optomotor responses of the body (OMR)
when exposed to visual stimuli simulating egomotion of the 
fish~\cite{BakSmiCoo15,KuboETAL14}. Both eye- and body movements 
generate space-variant patterns of local motion vectors on the retina 
which then have to be analyzed by subsequent processing stages. 
Neural algorithms
suggested for optic flow analysis usually consist of at least two
components, a local motion detector and a subsequent set of templates
or motion models for identifying typical patterns relating to
ego-motion maneuvers or encounters with obstacles and self-moving
objects such as prey or predator~\cite{FraChaKra04,Perrone92}.
Local motion detection can take place in the retina itself, as is
generally the case in lower vertebrates, or in early areas of visual
cortex. Higher brain areas analyzing optic flow patterns such as
the focus of expansion, rotational vertices, left or right yaw
rotations, etc., have been identified in mammalian MST 
cortex~\cite{Orban08} or in the zebrafish area pretectalis, APT~\cite{KuboETAL14}. 

Egomotion estimation from optic flow is 
subject to a large variety of established approaches derived from
geometric considerations~\cite{RauNeu12}. 
More recently, convolutional neural networks (CNNs) have 
shown remarkable characteristics, as they can learn
depth, motion fields and camera motion altogether 
in an unsupervised fashion~\cite{VijEtAl17,ZhouEtAl17}. Currently, CNN
architectures are state of the art for optic flow estimation~\cite{IlgEtAl17}
while other competitive approaches like~\cite{TimGoo15,WulBla15} exist that seek to 
estimate optic flow from a small (or sparse) number of matched templates.

In our model, local visual motion is encoded in the
direction-specific tuning curves of retinal ganglion cells and is not
subject to learning. Output from the retinal ganglion cells is then
fed into a layer of simulated APT-neurons which develop optic flow
analyzers.

\subsubsection*{Zebrafish visual system.}

Zebrafish retinal ganglion cells (RGCs), as well as pretectal cells,
exhibit clear tuning to the direction and orientation of drifting
gratings~\cite{AntSulMonHin16}. Movement direction is not covered
homogeneously, but clustered around three or four major visual field
directions~\cite{NikolaouETAL12}. The larval zebrafish retina contains
some 4000 ganglion cells with an average angular separation of about
2.5 degrees of visual angle.

RGCs project to APT, among other targets. The response characteristics
of APT neurons have been analyzed with visual stripe patterns (drifting
gratings) moving either forward or backward and presented to the left,
right, or both eyes~\cite{KuboETAL14}. Activity of {\em monocular}
neurons depends only on the stimulus delivered to one eye and can
therefore be considered to be directly driven from this eye's RGCs. In
contrast, {\em binocular} neurons combine input from both eyes to
generate specificities to forward or backward translation as well as to
clockwise and counter-clockwise rotation in the horizontal plane. 

\section{Visual Front End}

Realistic optic flow stimuli were generated from a virtual reality
simulation of observer motion in a fish tank, programmed in
{\em Blender}\footnote{https://www.blender.org}. The head of the
fish was modeled by two cameras rigidly moving together with a
rotation center somewhat behind the eyes.  The field of view was 160
by 160 degrees with a binocular overlap of $45$ degree (see
~\cite{KuboETAL14}). This results in central viewing directions of $\pm
57.5$ degree for the left and right eye.

\begin{figure}
\centering

\resizebox{0.953\textwidth}{!}{\includegraphics{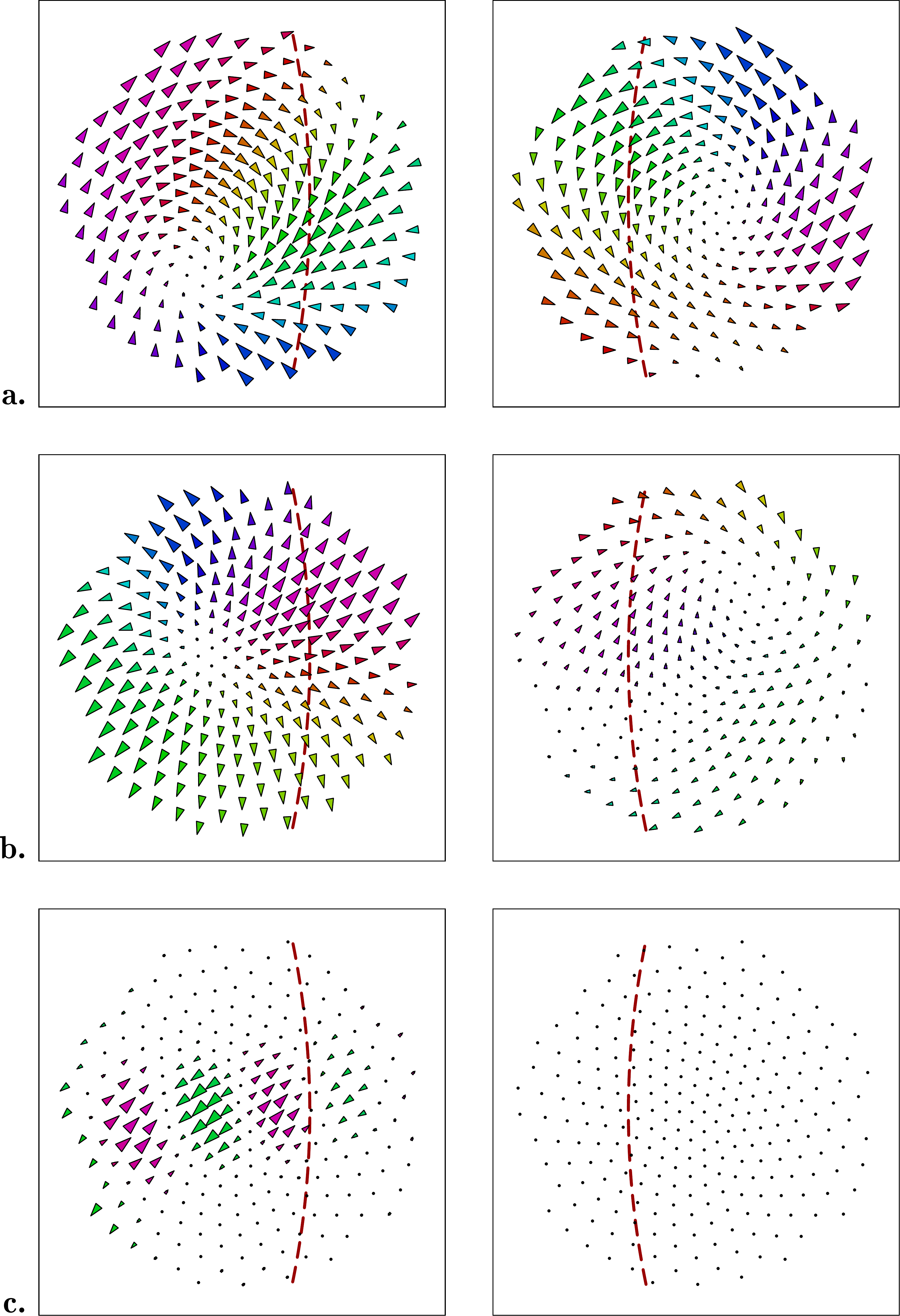}}

\caption{Sample binocular receptive fields from the sparse coding
  network. The red dotted lines mark the margin of binocular
  overlap. {\bf a.} Binocular whole-field neuron with spiral/rotatory
  characteristic. {\bf b.} Left-dominant whole-field neuron with
  elliptical focus of expansion in the left eye and a superposition of
  two curls in the right eye.  {\bf c.} Monocular Gabor-field}
\label{f:receptivefields}
\end{figure}

The virtual fish-tank contained objects at various distances from the
observer as well as objects in mid-water (floating plants and passing
fish) generating optic flow discontinuities in translational egomotion
(Fig.~\ref{fig:sim_examples}a,b).  Note that translatory optic flow
depends on object distance whereas rotatory optic flow does
not. Visibility was set either low (muddy water,
Fig.~\ref{fig:sim_examples}a) or high (clear water,
Fig.~\ref{fig:sim_examples}b). Overall, the scenery was built to
resemble the natural habitat of zebrafish as described in~\cite{SpeGerLawSmi08}.

Virtual fish were placed randomly in the environment and accelerated
by a short, random impulse both for translation and
rotation. Acceleration for all six degrees of freedom (DoF) were drawn
independently from a uniform, zero mean distribution, with an additional
scaling factor for the rotatory DoFs introduced in order to equalize
the average flow vector lengths of rotatory and translatory flow
components.  
After the acceleration impulse, the motion declined exponentially and
a two-frame motion sequence was recorded from the later (slower) parts
of this relaxation. Optic flow was calculated with FlowNet~2.0~\cite{IlgEtAl17}.

The fish retina was modeled as a spherical shell covering $160$
degrees in which 256 sampling points were placed using a simple
repellence algorithm (Fig.~\ref{fig:sim_examples}c). The planar camera
images were warped by stereographic projection and sampled at these
points. For each retinal sampling point $i$ the corresponding local
motion vector $(u_i,v_i)$ was represented by two signed variables
modeling the activity of pairs of RGCs tuned to opposite motion
directions (right/left, and up/down).

\section{LCA sparse coding}

For {\em unsupervised learning}, we used the locally competitive algorithm 
(LCA)~\cite{OlsFie96,RozJohBarOls08} which can be summarized as follows. Let
$\bm{x}= \{x_n\}_{n=1}^N$ denote the input signal, i.e.\ the output of
ganglion cells that encode local retinal motion. In sparse coding, the
goal is to reconstruct $\bm{x}$ as a linear combination $\bm{x}
\approx \sum_{k=1}^{K} a_k \bm{\varphi}_{k}$ with dictionary elements
\( \{ \bm{\varphi}_k \}_{k=1}^K \), and activation coefficients $\{
a_k \}_{k=1}^K $, for which sparsity is required~\cite{OlsFie96}. The
$\bm{\varphi}_k$ are vector fields from which the input vector field
can be reconstructed as a linear combination. According 
to~\cite{OlsFie97,RozJohBarOls08}, each $\bm{\varphi}_k$ can also be
considered as the receptive field of the $k$-th output neuron, if a
specific activation function with lateral feedback is assumed.  In our
application, the dictionary elements model the receptive fields of $K$
APT neurons. The vector $\bm{a} = \{a_k\}$ contains the coefficients
needed to reconstruct a given input pattern from the receptive
fields. In our simulations, we require $a_k \ge 0$ at all times. If we
write the $\bm{\varphi}_k$ as columns of a matrix $\bm{\Phi}$ we
obtain the error function $ E(\bm{a}, \bm{\Phi}) = \frac{1}{2} ~ \|
\bm{x} - \bm{\Phi} \bm{a} \|_2^2 \: + \: S(\bm{a})$, in which the
first term penalizes reconstruction errors and \(S(\bm{a})\) penalizes
non-sparse vectors \(\bm{a}\). While the original 
algorithm~\cite{OlsFie96} is based on the \(\ell^1\)-norm, i.e.\ the total
activity of \(\bm{a}\), the locally competitive algorithm (LCA) seeks
to minimize the \( \ell^0 \)-norm, i.e.\ the number of non-zero
$a$-values or the number of active units~\cite{RozJohBarOls08}. Since
$a_k \ge 0$, this amounts to choosing $ S(\bm{a}) = \sum_{k=1}^K
\lambda ~ {\cal H}(a_k-\lambda)$ where ${\cal H}$ is the Heaviside
function.

For the optimization algorithm see~\cite{OlsFie96,RozJohBarOls08}. The
algorithm was run in 
Petavision\footnote{{\tt https://petavision.github.io} and \protect\cite{schultzARxIV14}}
with $K = 512$ APT-neurons and $77,076$ motion fields each sampled at
256 retinal points for each eye ($N = 1024$). Examples of the
resulting $\bm{\varphi}_k$ are displayed as vector fields in
Figure~\ref{f:receptivefields}.

\section{Backpropagation}

For comparison, we also implemented a {\em supervised learning} 
version of the model that used the same
retinal encoding scheme and input data described above. Motion
sequences were labeled for egomotion by seven continuous variables,
three for the unit-vector of heading (translation), three for the unit
vector of the axis of rotation, and a non-negative one for rotational
speed. Note that translational speed cannot be recovered from optic
flow, so we did not attempt to teach this to the network. The network
contained three hidden layers with 1000, 600, and 200 units and an
output layer with seven units with the above encoding. Implementation
was carried out in TensorFlow\footnote{\tt https://tensorflow.org}.

The network was able to recover the heading direction with a mean
angular error of about $15$~degree and the axis of rotation with a mean
angular error of about $19$~degree.

\begin{figure}
\centering
\resizebox{0.955\textwidth}{!}{\includegraphics{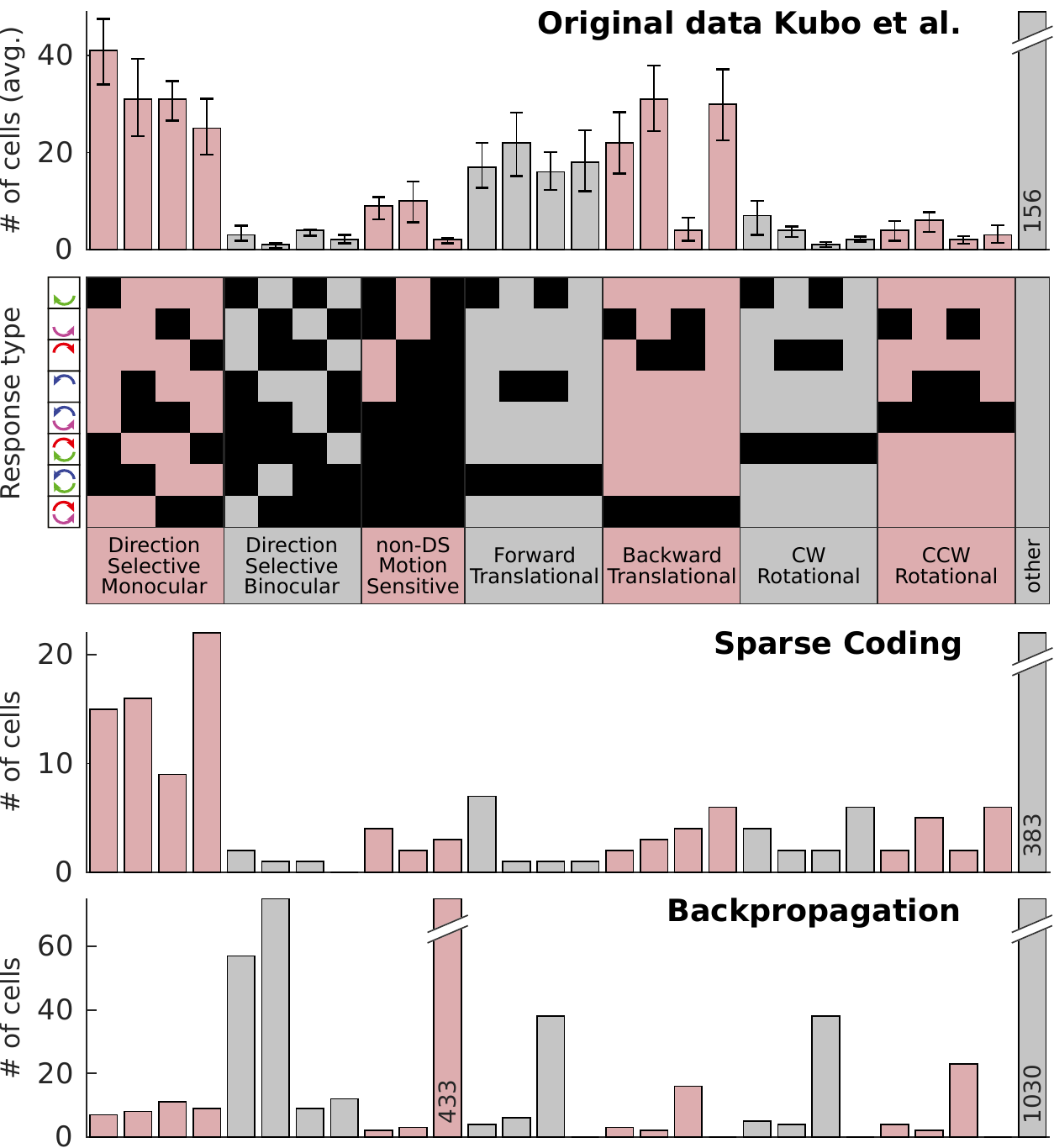}}

\caption{Summary of neuron response characteristics.  The top
  two panels are redrawn from~\protect\cite{KuboETAL14}. On the left
  of the ``{\bf Response type}'' panel, the little arrows symbolize
  optic flow stimulation when the fish is heading towards the left,
  i.e.\ the first row shows forward optic flow stimulation to the left
  eye, the second row backwards stimulation to the left eye and so
  on. The response types are indicated by the columns of black
  squares. E.g.\ the first column refers to neurons responding whenever
  there is forward stimulation to the left eye, irrespective of the
  stimulus delivered to the other eye, and so on. The histogram on top
  (``{\bf Original data}'') shows the frequency per fish of neurons of
  a given response type found in a sample of 3015 cells from six
  zebrafish larva APT. Most neurons are monocular direction selective
  (first block). Also, a substantial fraction of neurons specifically
  responding to global optic flow fields (forward translation etc.)
  was found. The third panel (``{\bf Sparse Coding}'') shows the
  results of the present study which are in good general agreement
  with the fish data. The ``{\bf Backpropagation}'' block shows the
  responses of the 1,800 units from all three hidden layers of the
  supervised learning network, which had been trained to classify optic
  flow patterns for egomotion.}
\label{f:results}
\end{figure}
\nocite{KuboETAL14}

\section{Results}

The simulations produce two types of data, i.e.\ models of vectorial receptive
fields, and neuronal responses to optic flow stimuli.   Receptive
fields will be discussed only for the sparse coding network since no
obvious interpretation was found for the backpropagation case. 

Figure~\ref{f:receptivefields} shows three typical examples out of the
set of 512 $\bm{\varphi}_k$ fields. Individual vector fields are
generally not realizable as optic flow fields in a rigid
environment. For example, Figure~\ref{f:receptivefields}a approximates a
pitch rotation (nose down) in both eyes, but the axes in the two eyes
are not properly aligned. Flow vectors are not purely tangential to
the pole but involve a spiral component. Figure~\ref{f:receptivefields}b
shows a left-dominant field with an expansion pattern in the left
eye. The focus is elongated as might be expected if two nearby foci
would superimpose. The right eye field is a superposition of two
rotational poles. We conjecture that ``dipole'' fields of this type
are needed to represent multiple axes of heading and rotation as
linear combinations of vector fields.  The two receptive fields of
Figure~\ref{f:receptivefields}a,b have high average $a_k$ values (rank 4
and 10 of the entire set). Figure~\ref{f:receptivefields}c shows a field
with low contribution to the reconstruction ($a_k$ rank 130) which is
representative of a large number of fields. It is monocular with
clearly delineated lobes of motion preferences in opposite directions,
resembling Gabor functions for the horizontal and vertical motion
components. Comparable, spatial frequency selective but non-localized 
fields were found by means of a PCA analysis by~\cite{WulBla15}. Together, these
findings mirror typical results when applied to images directly.

Binocular receptive fields obtained from either learning scheme were
further analyzed by calculating their response to spherical rotating
or translating grating stimuli as were used for receptive field
mapping in the zebrafish study by~\cite{KuboETAL14}.  Gratings can
move either forward or backward and can be presented to the left,
right, or both eyes. Altogether, four monocular and four binocular
stimulus types can be distinguished, see
Figure~\ref{f:results}. Each neuron or model neuron was classified for
its reaction to each of the eight stimulus types, resulting in $2^8 =
256$ response types. Of these, 27 optic-flow-related cases are shown
in Figure~\ref{f:results} both for the zebrafish recordings (upper
histogram) and for the two network simulations (lower
histograms). There is also a substantial number of cells not
classified into one of the illustrated 27 response types.

The response-type group ``direction selective monocular'' is most
frequent in the fish as well as in the sparse coding network, but not
in the backpropagation network. It includes neurons that react to the
stimulation of one eye, but ignore the stimulus of the other eye. On
their own, such neurons cannot analyze egomotion because they cannot
distinguish between forward translation and rotation to the
contralateral side. However, in the reconstruction approach of sparse
coding, they do seem to play an important role in describing the
binocular motion fields as well.

The next most frequent response type groups comprise binocular neurons
reacting to specific types of binocular optic flow such as translation
or rotation. The specificity of these responses is established by 
integrating directional information across both eyes.
Again, the sparse coding network seems to fit the data better
than the backpropagation network.

\section{Discussion}

In conclusion, receptive fields of zebrafish APT neurons are clearly
related to the statistics of environmental stimuli. The sparse coding
network seems to be closer to the data, but does not include a
mechanism of egomotion recovery. This recovery is implicit in the
backpropagation network, but the behavioral relevance of these
patterns is not guaranteed. In any case, more work is needed to
identify the detailed objective functions reflecting the information
requirements of the behaving fish.

Inspection of the vectorial receptive fields learned in the sparse
coding network (Fig.~\ref{f:receptivefields}) suggests that multiple
heading directions and axes of rotation are represented by base fields
that are not realizable as optic flow templates but provide a basis
for linear combination. This is in contrast to the coding by large
field templates in the fly~\cite{FraChaKra04} and the piecewise
construction of optic flow fields from local templates suggested for
mammals~\cite{Perrone92}.

\begin{sloppypar}
\printbibliography
\end{sloppypar}

\end{document}